\documentclass[aps,twocolumn,prd,epsf,showpacs,amsmath,amssymb]{revtex4}

\usepackage{graphicx}
\usepackage{dcolumn}
\usepackage{bm}

\newcommand{\be}{\begin{equation}}
\newcommand{\ee}{\end{equation}}
\newcommand{\ba}{\begin{eqnarray}}
\newcommand{\ea}{\end{eqnarray}}
\newcommand{\ban}{\begin{eqnarray*}}
\newcommand{\ean}{\end{eqnarray*}}

\newcommand{\la}[1]{\label{#1}}
\newcommand{\s}{\ensuremath{\psi(t,r)}}
\newcommand{\n}{\ensuremath{\nu(t,r)}}

\newcommand{\eq}[1]{(\ref{#1})}

\newcommand{\M}{\ensuremath{{\cal M}}}

\newcommand{\f}{\ensuremath{f(r)}}
\newcommand{\dw}{\ensuremath{{d\Omega^2}}}

\begin{document}

\title{Gravitational collapse of dustlike matter with heat flux}      

\author{Rituparno Goswami}

\affiliation{Theoretical Physics Institute, University of Alberta,
Edmonton, Alberta, Canada, T6G 2G7}

\email{goswami@phys.ualberta.ca}

\date{\today}

\begin{abstract}    
We present a new class of solutions to Einstein equations for 
the spherical collapse of dustlike matter coupled with heat flux. In this family of solutions
spacetime shear is necessarily non-zero. Also these solutions have an interesting property that
there is always a bounce before the singularity, which is caused entirely due to the 
dissipative processes. We show there exist open sets of initial data 
for which the bounce occurs before any trapped surface formation, 
making the star explode away to infinity. We also discuss the role of heat 
flow in generating spacetime shear and in 
modifying the effective inertial mass of the matter cloud.
\end{abstract}

\pacs{04.20.Dw, 04.70.-s, 04.70.Bw \hfill  
Alberta-Thy 07-07} 

\maketitle

\section{Introduction}

A considerable amount of work has continued in recent years
to model a realistic gravitational collapse scenario within the framework 
of Einstein's gravity 
(see e.g. 
~\cite{global},~\cite{gos1} 
and the references therein for some recent reviews).
The reason
for this interest is that general relativity generically admits the 
existence of spacetime singularities. These are extreme regions
in the spacetime where densities and spacetime curvatures typically 
blow up and the theory must breakdown.
\par
When a sufficiently massive star starts 
collapsing gravitationally on exhausting its nuclear fuel, it would 
not settle to a stable configuration such as a neutron star. What 
happens in such a case is an continual gravitational collapse ensues, where 
the sole governing force is gravity.
One of the interesting problems in this pursuit of study of gravitational 
collapse of such a massive star, is studying various dynamical solutions 
of Einstein's equations with radial heat flow in the interior, satisfying 
all the energy conditions. This radial heat flux is generated by 
various dissipative processes in the collapsing interior and indeed the 
relevance of these processes are central to a realistic astrophysical 
collapse.
\par
The earlier investigations done in this context, suggest that the final outcome of 
continual collapse in the presence of a radially outgoing heat flux 
is dramatically different from the usual Oppenheimer-Snyder picture 
in which a trapped surface develops before the singularity and ultimately 
the collapsing matter settle down to a black hole. For example, the solution 
presented in 
~\cite{las} 
describes the collapsing matter which radiates out mass due to the 
outward heat flow to an exterior Vaidya region 
(for details of the matching conditions please see
~\cite{san}) 
and through this process the complete star evaporates away leaving behind a 
flat Minkowski spacetime. 
A different solution was found in 
~\cite{ban1} and~\cite{ban2}, 
where the loss of matter due to heat flux stops 
the trapped surface formation and a naked singularity is formed as the 
end state of the gravitational collapse. 
Detailed investigations on the dissipative processes in the interior spacetime, 
using diffusion approximation was carried out in 
~\cite{herr1}-~\cite{herr5},
and it was shown that due to the {\it inertia of heat} (as described by Tolman 
~\cite{tol}), the collapse may slow down and stop before the singularity formation.  
\par
However, in all these earlier works we referred to above, the interior
spacetime was considered to 
be shearfree, which is rather a strong supposition. The basic aim of this paper is to 
investigate the final outcome for a more general class of collapsing
solutions of Einstein equations 
for a spherically symmetric spacetime, without the restriction of being shearfree.
For better clarity we take the matter to be dustlike ({\it i.e} 
the pressures are small 
compared to the density at all epochs) coupled to a radially outgoing heat flux 
and satisfying all the energy conditions. 
We discuss various regularity issues and do a near the center analysis 
of the class of solutions presented here, to show that in presence of heat flow 
the collapsing configuration always undergo a `bounce' before the singularity and hence 
the singularity formation is 
avoided. We also discuss the role of heat flow in generating spacetime shear and in 
modifying the effective inertial mass of the matter cloud.
    
\section{The collapsing system}

A general spherically symmetric metric in the coordinates 
$(t,r,\theta,\phi)$ can be written as,
\begin{equation}
ds^2=g_{ab}dx^adx^b+R^2(t,r)\dw
\label{eq:genmetric}
\end{equation}
where where $a,b$  run from $0$ to $1$, $x^0=t$, $x^1=r$ and $d\Omega^{2}$ 
is the line element on a two-sphere.
In general, we have freedom of coordinate transformations of the form 
$t'=f(t,r)$ and $r'=g(t,r)$. We use one transformation to make the $g_{tr}$ 
term vanish. So the metric now take the form,  
\begin{equation}
ds^2=-e^{2\n}dt^2+e^{2\s}dr^2+R^2(t,r)\dw
\label{eq:metric}
\end{equation}
The energy-momentum tensor for the dustlike matter coupled with heat 
flow is given as
\be
T^{ik}=\rho u^iu^k+q^iu^k+u^iq^k
\label{eq:emtensor}
\ee
where $\rho$ is the energy density, $u^i$ is the timelike unit vector 
(or the 4-velocity of the collapsing fluid) and  
$q^i$ is the heat flow vector.
We now use the other transformation to make the 3-velocity 
of the collapsing dust to vanish ({\it i.e.} the 4-velocity of the collapsing 
dust is of the form $u^i=e^{-\nu}\delta^i_t$), so that the frame is comoving
~\cite{landau}.
Now we do not have any gauge freedom of two variables left.
We still have the two scaling freedoms of one variable, which we would 
use later.
It is easily seen from \eq{eq:emtensor} that by redefining the density we can always 
make the heat flow to be orthogonal to the velocity vector such that $q^iu_i=0$.
In that case the vector $q^i$ would have the form,
\begin{equation}
q^i=Q(t,r)\delta^i_r\;.
\label{eq:q}
\end{equation}
The matter cloud is assumed to have a compact support with $0<r<r_b$, where $r_b$
denotes the boundary of the cloud, outside which it is to be 
matched through suitable junction conditions with an exterior 
spacetime geometry. For example, as we would like to model an astrophysical collapse 
scenario here, we would match the interior with an outgoing (exploding) generalized Vaidya 
exterior.
The function $R(t,r)$ is the area radius of a shell labeled `$r$' at an epoch `$t$'. 
And as we consider the situation where the cloud is collapsing initially we must have
$\dot R<0$ at the initial epoch. In a continual collapse scenario this
ultimately reaches $R=0$, which is the singularity where all
matter shells collapse to a zero physical radius.
We use the scaling freedom for the radial coordinate 
$r$ to write $R=r$ at the initial epoch $t=t_i$, and to differentiate the 
center of the cloud from the singularity 
(as in both these cases the area radius vanishes) 
introduce a function $v(t,r)$ as defined by
\begin{equation}
v(t,r)\equiv R/r \; .
\label{eq:R}
\end{equation}
We then have 
\begin{equation}
R(t,r)=rv(t,r),\; v(t_i,r)=1,\; v(t_s(r),r)=0 
\end{equation}
with $\dot{v}<0$.
The time $t=t_s(r)$ corresponds to the shell-focusing 
singularity at $R=0$, where the matter shell labeled `$r$' collapse to a 
vanishing physical radius.
We assume that the matter
field satisfies all the {\it energy conditions}, in terms of the density 
and heat flow vector this corresponds to ~\cite{ban1}
\begin{equation}
\rho\ge2|q| 
\la{eq:EC}
\end{equation}
where $|q|=g_{ik}q^iq^k$.
The dynamic evolution of the initial data is determined by the 
Einstein equations. For the metric (\ref{eq:metric}), and with the following definitions  
(to make the second order system of equations to a coupled first order system),
\begin{equation}
G(t,r)=e^{-2\psi}(R^{\prime})^{2}, H(t,r)=e^{-2\nu} (\dot{R})^{2}\;,
\la{eq:ein5}
\end{equation}
\begin{equation}
F=R(1-G+H)\;, 
\label{eq:ein4}
\end{equation}
we can write the independent Einstein's equations for the given matter field \eq{eq:emtensor}
as,
\begin{eqnarray}
F^{\prime}=\rho R^{2}R^{\prime}, & &
\dot{F}=0 \;, 
\label{eq:ein1}
\end{eqnarray}
\begin{equation}
\nu' R^2\rho=-e^{-(\psi+\nu)}\partial_t[e^{3\psi}QR^2]\;,
\label{eq:ein2}
\end{equation}
\begin{equation}
R'\frac{\dot{G}}{G}-2\dot{R}\nu'=Re^{(2\psi+\nu)}Q\; .
\label{eq:ein3}
\end{equation}
Here $(')$ denotes the partial derivative with respect to the co-ordinate `$r$' 
while $(\dot{})$ denotes partial derivative with respect to the co-ordinate `$t$'.
The function $F$ here has an 
interpretation of the mass function for the cloud, and it
gives the total mass in a shell of comoving radius 
`$r$' on any spacelike slice `$t=const$'. The energy conditions imply 
$F\ge0$. 
To preserve regularity at any non-singular epoch `$t$', we must have  
the mass function should vanish at the center of the cloud.
\par 
As in {\it Lemaitre Tolman Bondi}  
(LTB) models, we now try to find a new family of solutions with $\nu=\nu(t)$. 
This is an ansatz that we use here and this would only be consistent with some 
classes of heat flow and not the most general class.
Then we can always rescale the time 
co-ordinate to make the metric component $g_{tt}=-1$. We note that 
then we have no more gauge or scaling 
freedom left. Now from \eq{eq:ein1} we have, 
\be
F=F(r)=r^3\M(r)
\la{eq:F}
\ee
where $\M(r)$ is at least a $C^2$ function of the co-ordinate `$r$'.
It is to be noted that $F$ must have this form, which 
follows from the regularity 
and finiteness of the density profile at the initial epoch $t=t_i$, 
and at all other 
later regular epochs before the cloud collapses to the final
singularity at $R=0$. Also integrating \eq{eq:ein2} we have, 
\be
Q=\frac{r^3f(r)e^{-3\psi}}{R^2}
\la{eq:Q}
\ee
where $\f$ is another $C^2$ function, and the form of the function of 
integration is chosen in a way such that the heat flow vanishes at the
center of the cloud. We again would like to emphasize that this is not the 
most general class of the heat flux function. However as we would see later, 
near the center of the cloud and at any given {\it epoch} $v$, $Q\approx r\rho$. 
This definitely makes sense physically, as the heat flux must vanish at the 
center and also it should be increasing with increase in density.
Substituting the form of $\nu$ and \eq{eq:Q} in \eq{eq:ein3}, we have 
\be
R^\prime\frac{\dot{G}}{G}=\frac{e^{-\psi}r^3f(r)}{R}  
\ee
Multiplying both sides of the above equation by $-(1/2)(1/R'^2)$
and rearranging we get
\be
\partial_t[G^{(-1/2)}]=-\frac{1}{2}\frac{r^3\f}{R(R')^2}\;.
\la{eq:G1}
\ee
Now to construct classes of solutions 
to Einstein equations, let us define a suitably 
differentiable function $A(r,v)$ as follows,
\begin{equation}
-\frac{1}{2}\frac{r^3\f}{R(R')^2}=A(r,v)_{,v}\dot{v}
\label{eq:A}
\end{equation}
We would like to emphasize here that this is no special choice but merely a 
construction which would help us decouple the Einstein equations and 
find classes of solutions. Using this definition we can now integrate \eq{eq:G1} 
to get
\be
G=\frac{1}{[A(r,v)+r^2b(r)]^2}\;.
\la{eq:G2}
\ee
Here again, the function $b(r)$ is a suitably differentiable function
of integration. We also would 
like to point out here that the function $A(r,v)=A(r)$ corresponds to $\f=0$ 
(that is the situation where $Q=0$) and we get back the well known LTB solution.
Now from \eq{eq:ein4} we see that in order to make $\dot{v}$ well defined at the 
center of the cloud ($r=0$), we should have 
$[A(r,v)+r^2b(r)]^{(-2)}\equiv 1+r^2B(r,v)$, 
where $B(r,v)$ is another suitably differentiable function. Or in other 
words the function $A(r,v)$ must have the form,
\be
A(r,v)=\frac{1}{\sqrt{1+r^2B(r,v)}}-r^2b(r)
\la{eq:Aform}
\ee
We note that the above equation is not really any ansatz 
or a special choice, but quite a generic class of function
consistent with and allowed by the required regularity 
conditions of metric functions at the center of the cloud.
Comparing \eq{eq:G2} with \eq{eq:ein5} and using \eq{eq:Aform} we get
\be
g_{rr}=e^{2\psi}= \frac{(R)'^2}{1+r^2B(r,v)}\;.
\la{eq:grr}
\ee
Using the above equation and \eq{eq:ein4} we have,
\be
\dot{v}=-\sqrt{B(r,v)+\frac{\M(r)}{v}}.
\la{eq:vdot}
\ee
The negative sign in the above equations corresponds 
to a initial collapse scenario, where we have $\dot{R}<0$ initially.
Also using equation \eq{eq:Aform} in \eq{eq:A} we get,
\be
v+rv'=\left[-\frac{\f(1+r^2B(r,v))^{(3/2)}}
{vB(r,v)_{,v}\sqrt{B(r,v)+\frac{\M(r)}{v}}}\right]^{(1/2)}
\la{eq:vdash}
\ee
Now from equations \eq{eq:vdot} and \eq{eq:vdash} we get the expressions of 
$v'$ and $\dot{v}$ in terms of $r$, $v$ and the function $B(r,v)$ and it's partial 
derivative with respect to $v$.
Thus, to get a solution of $v(t,r)$ we need the {\it Pfaffian
differential equation}, $ v'dr+\dot{v}dt-dv=0$ , to be integrable.
This integrability conditions gives us the required second order differential
equation for the function $B(r,v)$.
To show that solutions of this equation exist, it is sufficient to
find one solution. We have already seen that $B=B(r)$ is a solution which 
corresponds to the LTB solution with $\f=0$.
In that case, providing the value $B(r,1)$, completely determine the allowed function
$B(r,v)$. From \eq{eq:vdash} it is clear that we get $B(r,v)_{,v}|_{v=1}$ (which corresponds 
to providing $Q_{,t}(t_i,r)$) from the other three free functions.
\par
Thus, we see that in this class of models there are three free functions of the 
co-ordinate `$r$', namely $\M(r)$, $B(r,1)$ and $\f$.
The first function gives the density profile, the second one describes the velocity profile
of the infalling shells 
while the third describes the flux profile at the initial epoch. 
Given these initial profiles we can solve the system of 
Einstein equations completely, and from \eq{eq:vdot} we can write the
solution of $v(t,r)$ in the integral form as,
\be
t(v,r)=\int_v^1\frac{\sqrt{v}dv}{\sqrt{vB(r,v)+\M(r)}}
\la{eq:sol} 
\ee
We note that the variable `$r$' is treated as a constant in the above equation.
The above equation gives the time taken for a shell labeled `$r$' to 
reach a particular {\it epoch} `$v$' from the initial epoch $v=1$.
\par
To complete a collapse model, we need to match the
interior spacetime to a suitable exterior spacetime. As we are
interested here in modeling collapse of astrophysical objects (such as 
massive stars), we have assumed the matter to have compact support
at the initial surface, with the boundary of the cloud being 
at some $r=r_b$. We match the interior with a general class of exterior metrics, 
which are the generalized Vaidya spacetimes (it describes the matter  
which is a combination of matter fields of 
{\it Type I} and {\it Type II} ~\cite{wang}).   
at the boundary hypersurface $\Sigma$ given by $r=r_b$. 
For the required matching we use the {\it Israel-Darmois} conditions 
(\cite{match4}-~\cite{match6}), 
where we match the first and second fundamental forms 
(the metric coefficients
and the extrinsic curvature respectively), at the boundary
of the cloud. The
metric in the exterior of $\Sigma$ is given by,
\begin{equation}
ds^2_{+}=-\left(1-\frac{2M(r_v,V)}{r_v}\right)dV^2-2dVdr_v+r_v^2d\Omega^2
\label{eq:metricvaidya}
\end{equation} 
where $V$ is the retarded (exploding) null co-ordinate and $r_v$
is the Vaidya radius. For details of matching conditions we refer to
~\cite{gos1}.

\section{Analysis of the solution near the center}

To understand the dynamics of the collapsing system in more transparent way, 
let us consider a small neighborhood of the center of the cloud. That is we consider 
the value of the coordinate `$r$' to be small enough so that we can neglect the 
expressions containing the higher powers of $r$. Let us consider the free functions 
having the following Taylor expanded form near the center,
\ba
f(r)=f_0+f_2r^2+\cdots \nonumber \\
\M(r)=M_0+M_2r^2+\cdots 
\ea
We also write the function $B(r,v)$ in the form,
\be
B(r,v)=B_0(v)+B_2(v)r^2+\cdots
\ee
We know that near the center $R'\approx v$. 
Putting the above equations in \eq{eq:vdash}
and neglecting higher powers of $r$, we get,
\be
B_0(v)_{,v}\sqrt{B_0(v)+\frac{M_0}{v}}=-\frac{f_0}{v^3}
\ee
Now since we consider the heat flux is {\it outwardly} directed we must have 
$f(r)<0$ for all $r$ and hence $f_0<0$~\cite{ban1}. Hence we get the following ordinary 
differential equation for the evolution function $B_0(v)$,
\be
B_0(v)_{,v}=\frac{|f_0|}{v^{(5/2)}\sqrt{B_0(v)v+M_0}}
\la{eq:B}
\ee
\begin{figure}[tb]
\hspace*{0.1cm}
\includegraphics[width=6cm, height=5cm]{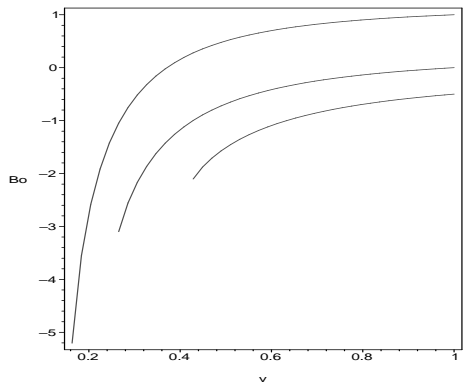}
\caption{The function $B_0(v)$ as found by numerical integration, for different initial values.
The constants were taken as $|f_0|=0.5$, $M_0=1$
We can see that even if we start from a positive value at the initial epoch, the function
becomes negative at the later epoch causing a bounce.}
\end{figure}

It is not possible to integrate the above equation analytically in a 
closed form. However numerical integration shows that even if we 
choose $B_0(1)>0$ at the initial epoch, the function $B_0(v)$ changes 
sign at some later epoch and there always exist an epoch $v_b\in(0,1]$
where $\dot{v}=0$. In other words the collapse of the small neighborhood of 
the central shell 
{\it stops} and shells bounces back before the spacetime singularity.
Figure 1 shows the function $B_0(v)$ as a function of $v$ and we see that there 
is an epoch at which $B_0(v)_{,v}$ blows up. From the above equation it is clear that 
it is the same epoch in which $\sqrt{B_0(v)v+M_0}\rightarrow 0$ or the quantity 
$\dot{v}=0$ (as shown in Figure 2) and the shells rebounce. Though this 
analysis is for a small neighborhood around the center, we note from equation 
\eq{eq:vdash} that we are only considering those solutions where $R'>0$ throughout 
the spacetime. In other words we are considering solutions free of any shell 
crossing singularities. In this scenario, if there is a bounce around the center, 
we can conclude the the whole collapsing matter bounces back and there
are no spacetime singularities.
\par
From equation \eq{eq:EC} we see that, as $|q|\approx r^2$ near the center, 
all energy conditions are satisfied in the neighborhood of the
center. Further to this, 
we can always choose the free functions $F(r)$, $\f$ and $B(r,1)$ in such a way 
that energy conditions are obeyed throughout the spacetime. To inspect more 
carefully the reason for this bounce, in spite of satisfying 
all energy conditions, let us consider the `{\it acceleration}' term, 
\be
\ddot{v}=\frac{1}{2}\left[\frac{|f_0|}{v^{(5/2)}\sqrt{B_0(v)v+M_0}}-
\frac{M_0}{v^2}\right]  
\la{eq:accn}
\ee

\begin{figure}[tb]
\hspace*{0.1cm}
\includegraphics[width=6.5cm, height=5cm]{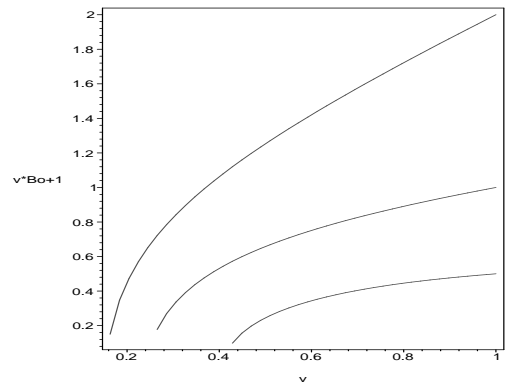}
\caption{The plot of the $v\dot{v}^2$ as a function of $v$ for different initial values. 
The constants were taken as $|f_0|=0.5$, $M_0=1$. 
It is seen that before the spacetime singularity $v=0$, the collapsing velocity 
goes to zero causing a bounce.}
\end{figure}

From the above equation it is clear that even if we choose the initial data in such a 
way that at the initial epoch $\ddot{v}<0$ ({\it i.e.} the collapse is accelerating), 
as the collapse proceeds the first term in the equation starts 
dominating and ultimately 
there is a {\it decelerating effect} as $\ddot{v}>0$. 
At the epoch of the bounce we have 
$\ddot{v}\rightarrow\infty$, that is, there is an infinite deceleration. 
\par
There is also an alternative thermodynamical explanation to this 
phenomenon. As shown in 
~\cite{herr5},
if we couple the Einstein equations with {\it Israel Stewart}
transport equation 
~\cite{israel}
we can easily see that the {\it Effective Inertial 
Mass Density} of the infalling matter (as defined by the `Newtonian' form of 
the equation of motion) reduces due to the presence of heat flow and
becomes negative as the collapse proceeds. For the dustlike matter
field we consider here, the effective inertial mass density behaves
like $\rho(1-\alpha)$, where the function $\alpha$ depends on
the temperature, density, thermal conductivity and relaxation
time. During the collapse the temperature of the cloud may increase 
enormously which in turn can make $\alpha>1$. 
Thus in spite of the fact that infalling matter should gain momentum 
due to the outgoing flux, 
the negative effective inertial mass density actually creates a `force' against 
the motion and the collapse stops before the singularity.
\par
Coupling of heat flux with collapsing matter also has another
important physical effect. We know, for the given metric \eq{eq:metric},
the shear tensor for the collapsing matter is given by,
\begin{equation}
\sigma_\phi^\phi=\sigma_\theta^\theta=-\frac{1}{2}\sigma_r^r=\frac{1}{3}
e^{-\nu}\left(\frac{\dot{R}}{R}-\dot{\psi}\right)
\la{eq:shear}
\end{equation}
Now, in the absence of any heat flow, if we consider $B(r,v)=1$ and 
$\M(r)=M_0$, then the spacetime is shearfree and the collapse becomes 
Oppenheimer Snyder homogeneous dust collapse. However if we couple the
matter with heat flux then we see that the shear tensor is necessarily 
non-zero. In other words even if we start with homogeneous matter and 
couple it with heat flux then the matter becomes inhomogeneous at 
later epochs and spacetime shear develops. We know that presence
of shear delays the formation of trapped surface 
~\cite{JDM}.
Also the loss of matter due to continuous radiation in the generalized 
Vaidya exterior further helps in delaying the trapped surface
formation. In that case we can easily see that there exist a large 
class of collapsing solutions (depending on choice of initial data and) 
for which the collapsing star bounces 
back before any trapped region forms, that is, the star explodes away
to infinity.

\section{Concluding Remarks}

In this paper we presented a new family of solutions which generalizes
the well known Lemaitre-Tolman-Bondi solutions, for the collapse of dustlike 
matter with heatflux and the spacetime shear is necessarily non-zero. The most 
important feature which emerges from these class of solutions 
is that, there is always a bounce before the singularity, which is 
caused entirely due to the dissipative processes. Also we showed, 
there exist open sets of initial data for which the bounce 
occurs before any trapped surface formation, making the star explode 
away to infinity. 
\par
Such a study of gravitational collapse is warranted for several
reasons. First of all, in the later stages of a massive star evolution, 
due to considerable decrease in opacity of fluids (for photons and neutrinos), 
heat conduction does play a very important role and that cannot
possibly be neglected. It is shown, for example, that this phenomenon
of
decreasing opacity give rise to neutrino radiative heat conduction
~\cite{arnett}
in the stellar fluid which may increase the temperature of the
collapsing star to a large extent. Secondly, as we have already seen, 
the general properties of the collapse in presence of heat flux 
is dramatically different from when it is absent and we should
give it a careful consideration, because there may exist several
alternate pathways through which a massive star may actually die. 
And finally as we know there are deep paradoxes associated with 
spacetime singularities where the classical theory breaks down. 
While one expects a successful quantum picture of spacetime 
would actually be able to resolve these singularities, it is always 
welcome if they are removed at the classical level.

\noindent  

\section*{Acknowledgments}  
\noindent
The author would like to thank Prof. Valeri Frolov for extremely helpful 
discussions throughout the project. Also special thanks to Prof. Pankaj S Joshi, 
Prof. Naresh Dadhich and Prof. Louis Herrera.  
The research was supported by the Natural
Sciences and Engineering Research Council of Canada.

\end{document}